\documentclass[12pt,fleqn]{article}
\usepackage{psfig}
\begin{document}
\begin{center}
{\Large{\bf Recent progress on the chiral 
unitary approach to  meson meson and meson baryon interactions}}
\end{center}

\vspace{.7cm}

{\large{ E. Oset$^ {1,2}$, J. A. Oller$^ {2}$,  J. R. Pel\'aez$^ {3}$, 
A. Ramos$^ {4}$, H. C. Chiang$^ {5}$, F. Guerrero$^ {2}$, S. Hirenzaki$^ {6}$,
 T. S. H. Lee$^ {7}$, E. Marco$^ {1,2}$, J. C. Nacher$^{1,2}$, Y. Okumura$^{6}$,
 A. Parre\~no$^ {8}$, H. Toki$^{1}$,
M. Vicente-Vacas$^ {2}$ }}

\vspace{0.8cm}

{\it $^1$ Research Center for Nuclear Physics (RCNP), Osaka University, Ibaraki,
 Osaka 567-0047, Japan.}

{\it $^2$  Departamento de F\'{\i}sica Te\'orica and IFIC, 
Centro Mixto Universidad de Valencia-CSIC
46100 Burjassot (Valencia), Spain.}

{\it $^3$ Departamento de F\'{\i}sica Te\'orica, Universidad Complutense,
Madrid, Spain.}

{\it $^4$ Departament d'Estructura i Constituents de la Materia, Universitat de
Barcelona, Diagonal 647, 08028 Barcelona, Spain.}

{\it $^5$ Institute of High Energy Physics, Academia Sinica, Beijing, China.}

{\it $^6$ Physics Department, Nara Women's University, Japan.}

{\it $^7$ Argonne National Laboratory, Argonne, U.S.A.}

{\it $^8$ Institute of Nuclear Theory, University of Washington, Seattle, U.S.A.}

\vspace{0.6cm}

\begin{abstract}
{\small{We report on recent progress on the 
chiral unitary approach, analogous to the effective range expansion
in Quantum Mechanics, which is shown to have a much larger convergence radius
than ordinary chiral perturbation theory, allowing one to reproduce data for
meson meson interaction up to 1.2 GeV. Applications to physical processes so
far unsuited for a standard chiral perturbative approach are presented. Results
for the extension of these ideas to the meson baryon sector are discussed,
together with applications to kaons in a nuclear medium and $K^-$ atoms.}} 
\end{abstract}
\vspace{.5cm}

\begin{center}
{\bf \large

Invited Talk Presented at the KEK Tanashi Symposium on 

Physics of Hadrons and Nuclei

Tokyo, December 1998

In honor of Prof. K. Yazaki.}
\end{center}

\vspace{0.5cm}
\section{Chiral Unitary Approach}

Chiral perturbation theory ($\chi PT$) has proved to be a  very suitable
instrument to implement the basic dynamics and symmetries of the meson meson
and meson baryon interaction [1] at low energies. The essence of the
perturbative technique, however, precludes the possibility of tackling problems
where resonances appear, hence limiting tremendously the realm of
applicability. The method that we expose naturally leads to low lying
resonances and allows one to face many problems so far intractable within
$\chi PT$.

The method incorporates new elements: 1) Unitarity is implemented exactly;
2) It can deal with coupled channels allowed with pairs of particles from the 
octets of stable
pseudoscalar mesons and ($\frac{1}{2}^+$) baryons; 3) A chiral
expansion in powers of the external four-momentum of the lightests 
pseudoscalars
is done for
Re $T^{-1}$, instead of the $T$ matrix itself which is the case in standard $\chi
PT$.

We sketch here the steps involved in this expansion for the meson meson
interaction. One starts from a $K$ matrix approach in coupled channels where
unitarity is automatically fulfilled and writes 
\begin{equation}
T^{-1} = K^{-1} - i\,\sigma ,
\end{equation} 
where $T$ is the scattering matrix, $K$ a real matrix in the physical 
region
and $\sigma$
is a diagonal matrix which measures the phase-space available for the intermediate
states
\begin{equation} 
\sigma_{nn}(s) = - \frac{k_n}{8\pi\sqrt{s}}\,\theta\left(s - (m_{1n} + m_{2n})^2\right),
\end{equation}
where $k_n$ is the on shell CM momentum of the meson in the intermediate state
$n$ and $m_{1n}$, $m_{2n}$ are the masses of the two mesons in the state $n$. The meson
 meson states considered here are $K\bar{K}$, $\pi\pi$, $\pi\eta$, $\eta\eta$,
 $\pi K$, $\pi\bar{K}$, $\eta K$, $\eta\bar{K}$. Since $K$ is real, from eq. (1)
 one sees that $K^{-1}$ = Re $T^{-1}$. In non-relativistic
 Quantum Mechanics, in the scattering
 of a particle from a potential, it is possible to expand
 $K^{-1}$ in powers of the momentum
 of the particle at low energies as follows (in the s-wave for simplicity)
\begin{equation} 
\hbox{Re}\,T^{-1}\equiv K^{-1} = \sigma\cdot ctg\delta\, \propto -\frac{1}{a} 
+ \frac{1}{2}r_0 k^2 ,
\end{equation}
with $k$ the particle momentum, $a$ the scattering length and $r_0$ the effective
range.

The ordinary  $\chi$PT expansion up to $O(p^4)$ is given by [1]
\begin{equation}
T = T_2 + T_4 ,
\end{equation}
where $T_2$, which is $O(p^2)$, is obtained from the lowest order chiral
Lagrangian, $L^{(2)}$, whereas $T_4$ contains one loop diagrams in the s, t, u
channels, constructed from the lowest order Lagrangian, tadpoles and the
finite contribution from the 
tree level diagrams of the $L^{(4)}$ Lagrangian. This last contribution, 
after a suitable renormalization, is just a polynomial, $T^{(p)}$.
Our $T$ matrix, starting from eq. (1) is given by
\begin{equation}
T = [\hbox{Re}\, T^{-1} - i\,\sigma]^{-1}
\equiv T_2 \,[T_2 \,\hbox{Re}\, T^{-1}\, T_2 - i\,T_2\,\sigma\, T_2]^{-1}\, T_2 ,
\end{equation}
where, in the last step, we have multiplied by $T_2 T_2^{-1}$ on the left and 
$T_2^{-1} T_2$ on the
right for technical reasons. But using standard $\chi$PT we obtain the following
expansion up to order $O(p^4)$,
\begin{equation}
T_2 \,Re T^{-1}\, T_2 = T_2 - \hbox{Re}\, T_4 ...
\end{equation}
and hence, recalling that $\hbox{Im}\, T_4 = T_2\,\sigma \,T_2$, one obtains
\begin{equation}
T = T_2\,[ T_2 - T_4]^{-1}\, T_2,
\end{equation}
which is the coupled channel generalization of the inverse amplitude method of
[2].

        Once this point is reached one has several options to proceed:

a) A full calculation of $T_4$ within the same renormalization scheme as in
$\chi PT$ can be done. The eight $L_i$ coefficients from $L^{(4)}$ are then fitted
to the existing meson meson data on phase shifts and inelasticities up to 1.2 GeV, where
4 meson states are still unimportant. This procedure has been carried out in
[2,3]. The resulting $L_i$ parameters are compatible with those used in $\chi PT$. 
At low energies the $O(p^4)$ expansion for $T$ of eq. (7) is identical to that
in $\chi PT$. However, at higher energies the nonperturbative structure of eq. (7),
 which implements unitarity exactly, allows one to extend the information
 contained in the chiral Lagrangians to much higher energy than in ordinary 
$\chi$ PT, which is up to about $\sqrt{s}\simeq 400 $ MeV. Indeed it
reproduces the resonances present in the L = 0, 1 partial waves.

\vskip .2cm

b) A technically simpler and equally successful additional approximation
 is generated by ignoring the crossed channel loops and
tadpoles and reabsorbing them in the $L_i$
coefficients given the weak structure of these terms in the physical region.
The fit to the data with the new $\hat{L}_i$ coefficients reproduces the whole meson
meson sector, with the position, widths and partial decay widths of the
$f_0(980)$, $a_0(980)$, $\kappa(900)$, $\rho(770)$, $K^\ast(900)$ resonances in good
agreement with experiment [4]. A cut off regularization is used in [4] for the
loops in the s-channel. By taking the loop function with two intermediate
mesons
\begin{equation}
G_{nn}(s) = i\int\frac{d^4 q}{(2\pi)^4}\, \frac{1}{q^2 - m_{1n}^2 + i\epsilon}
\, \frac{1}{(P-q)^2 - m_{2n}^2 + i\epsilon},
\end{equation}
where $P$ is the total meson meson momentum, one immediately notices that
\begin{equation}
\hbox{Im}\, G_{nn}(s) = \sigma_{nn}.
\end{equation}
Hence, we can write
\begin{equation}
\hbox{Re}\, T_4 = T_2\, \hbox{Re}\, G\, T_2 + T_4^{(p)},
\end{equation}
where $\hbox{Re}\, G$ depends on the cut off chosen for $|\vec{q}|$. This means that the
$\hat{L}_i$ coefficients of $T_4^{(p)}$ depend on the cut off choice, much as the
$L_i$ coefficients in $\chi PT$ depend upon the regularization scale.

\vskip .2cm

c) For the L = 0 sector (also in L = 0, S = $-1$ in the meson baryon interaction)
a further technical simplification is possible. In these cases it is possible
to choose  the cut off such that, given the relation between $\hbox{Re}\, G$
and $T_4^{(p)}$, this latter term is very well approximated by 
$\hbox{Re} T_4= T_2 \,\hbox{Re}\, G\, T_2$. This is impossible in those cases
because of the predominant role played by the unitarization of the lowest
order $\chi PT$ amplitude, which by itself leads to the low lying resonances,
 and because other genuine QCD resonances appear at higher energies.

 In such a case eq. (5) becomes 
\begin{equation}
T = T_2 \,[T_2 - T_2\, G \,T_2]^{-1} \,T_2 = [1 - T_2 \,G]^{-1}\, T_2,
\end{equation}
or, equivalently,
\begin{equation}
T = T_2   + T_2 \,G \,T,
\end{equation}
which is a Bethe-Salpeter equation with $T_2$ and $T$ factorized on shell outside
the loop integral, with $T_2$ playing the role of the potential. This option has
proved to be successful in the L = 0 meson meson sector in [5] and in the
L = 0, S = $-1$ meson baryon sector in [6].

        In the meson baryon sector with S = 0, given the disparity of the
        masses in the coupled channels $\pi N$, $\eta N$, $K\Sigma$,
        $K\Lambda$,
        the simple ``one cut off approach'' is not possible. In [7] higher
        order Lagrangians are introduced while in [8] different subtraction
        constants (or equivalently different cut offs) in G are incorporated
        in each of the former channels leading in both cases to acceptable
        solutions when compared with the data.

\section{Applications for processes involving pairs of mesons.}

Given the shortness of space we shall not show results on the meson meson
and meson baryon scattering that can be found in [4, 6], together with all
other technical details. Instead, we make a short summary of applications of
these ideas to other processes with results for the latest developments.

One of the applications in the meson meson sector is the study of the $\gamma
\gamma\rightarrow\pi^+\pi^-$, $\pi^0\pi^0$, $K^+K^-$, $K^0\bar{K^0}$, $\pi\eta$ 
reactions. The $\gamma \gamma\rightarrow\pi^+\pi^-$, $\pi^0\pi^0$ reaction
has been one of the standard places to test $\chi PT$ [9], with the obvious
limitations to small energies. The new techniques have allowed to extend
the calculations up to about $\sqrt s = 1.4$ GeV  and include the $K^+K^-$,
 $K^0\bar{K^0}$, $\pi\eta$ channels that were not accessible with $\chi PT$.
  Results for all these channels are presented in [10] where a good agreement
  with experiment is found in all cases.

  The decay channels of the $\phi(1020)$
  resonance also offer a good testing ground for the chiral unitarity theory.
   In [11] the $\phi\rightarrow\gamma K^0\bar{K^0}$ decay channel was studied
   providing a calculation of the background in future experiments testing CP violation at
   $DA\Phi NE$. In a recent work these ideas have been extended to study the
   $\rho^0\rightarrow\pi^+\pi^-\gamma$, $\pi^0\pi^0\gamma$ and 
$\phi\rightarrow\pi^+\pi^-\gamma$, $\pi^0\pi^0\gamma$ process [12]. The latter
   proceeds via $K^+K^-$ loops as shown in fig. 1.
\vspace{1.cm}
\begin{figure}[h]
\centerline{\protect
\hbox{
\psfig{file=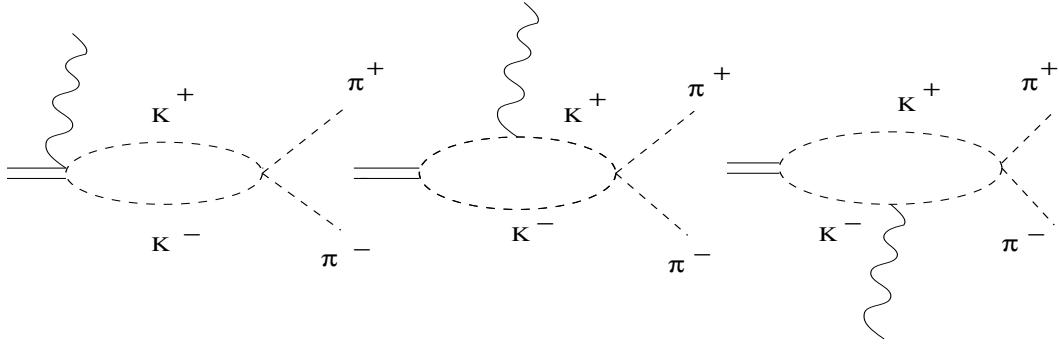,height=4.5cm,width=14.0cm,angle=-90}}}
\caption{Diagrams contributing to $\phi\rightarrow\pi^+\pi^-\gamma$ decay.}
\label{Fig.1}
\end{figure}

  The transition amplitude for this process behaves as
\begin{equation}   
t^\gamma\,  \propto\, e\, g_\phi\tilde{G}(M_ \Phi, M_I)
\, t_{K^+ K^-\rightarrow\pi^+\pi^-}(M_I),
\end{equation} 
where $g_\phi$ is the $\phi\rightarrow K^+K^-$ decay coupling, $M_I$ the
invariant mass of the $\pi^+\pi^-$ system and $\tilde{G}$ sums the loop functions
of the three diagrams, which using arguments of gauge invariance one can prove
to be finite. At the same time $t_{K^+K^-\rightarrow\pi^+\pi^-}(M_I)$, as
shown in [11], is the on shell strong scattering matrix for $K^+ K^-\rightarrow
\pi^+\pi^-$, which is evaluated using the chiral unitary techniques in [12].

\vspace{0.7cm}
\begin{figure}[h]
\centerline{\protect
\hbox{
\psfig{file=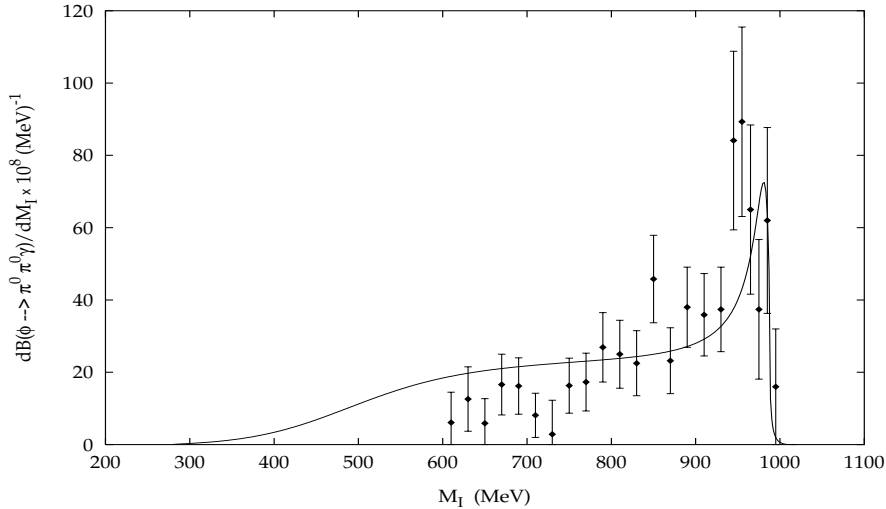,height=7cm,width=11.5cm,angle=-90}}}
\caption{${d\, B}/{d\,M_I}$ for $\phi\rightarrow\pi^0\pi^0\gamma$, with
 $M_I$ the invariant mass of the $\pi^0\pi^0$ system. Experimental points from
 [14]. Only statistical errors are shown.}
\label{Fig.2}
\end{figure}

The
branching ratio $\Gamma_{\pi^+\pi^-\gamma}/{\Gamma_{\phi}}$ obtained is
2.6 $10^{-4}$ while the $f_0$ peak contributes 0.76 $10^{-4}$. This
latter value is compatible and close to the latest boundaries of $B< 1-7 \cdot 10^{-4}$ 
obtained in Novosibirsk [13]. After completion of this work a publication has appeared
\cite{acha},
 where the spectrum for $\phi\rightarrow\pi^0\pi^0\gamma$ is measured. Taking
 into account that the rate for this process is one half of the one for
$\phi\rightarrow\pi^+\pi^-\gamma$ , the value for the branching ratio
which are obtain is 1.3 $10^{-4}$ which compares very well with the
experimental value (1.14 $\pm$ 0.10 $\pm$ 0.12) $10^{-4}$. The invariant
mass distribution is also in agreement with the experimental one, within
statistical and systematic errors, and this latter distribution shows
clearly the $f_0$ peak as predicted by theory. In fig. 2 we show our predicted
results compared with the recent measurements of [14]. The agreement found with
the whole spectrum is a good support for the chiral unitary theory used.

Another recent calculation is the reaction $\gamma
p\rightarrow f_0(a_0) p$ close to threshold [15], which requires energies
of $E_\gamma\simeq 1.7$ GeV, or larger, and which can be studied in planned
experiments at SPring8/RCNP [16] or TJNAF.

The scattering of mesons in a nuclear medium is an interesting problem
and predictions of a quasibound $\pi\pi$ system just below threshold have been
made [17]. Interestingly, an enhancement in the mass distribution of two pions
close to threshold in the $\pi^- A\rightarrow\pi^+\pi^-A^{\prime}$ reaction is
seen [18] which is absent in the $\pi^+ A\rightarrow\pi^+\pi^+A^{\prime}$. The
calculations done in [19] using the chiral unitary model and medium
modifications lead to an appreciable enhancement of the strength around the two pion
threshold, although peaks corresponding to the bound state do not appear.
The results are remarkably similar to those obtained using a meson exchange
picture for the $\pi\pi$ interaction, yet imposing minimal chiral constraints
[20], hence stressing the role of chiral symmetry in this process.

Another interesting development along these lines is the work of [21], where,
by taking the lowest order chiral Lagrangian and assuming that higher orders
of the chiral Lagrangian are generated by the exchange of genuine resonances
which survive in the large $N_c$ limit [22], the unitary approach together
with the comparison with data, allows one to distinguish between the genuine
resonances corresponding to QCD states, surviving in the large $N_c$ limit,
and scattering resonances which appear only as a consequence of the nature of
the lowest order Lagrangian together with unitarity.

These are only examples of problems which can be tackled with the new
techniques. Given the broad range of applications of $\chi PT$ and the
increased range of energies now accessible with the chiral unitary approach, one
can envisage a fruitful field of applications with the new approach in strong, weak and
electromagnetic processes at intermediate energies.

\section{Applications in the meson baryon sector}

As quoted above, a good description of the $K^-p$ and coupled channel interaction 
is obtained in terms of the lowest order Lagrangians and the Bethe Salpeter
equation with a single cut off. One of the interesting features of the approach
is the dynamical generation of the $\Lambda(1405)$ resonance just below the
$K^-p$ threshold. The threshold behavior of the $K^-p$ amplitude is thus
very much tied to the properties of this resonance. Modifications of these
properties in a nuclear medium can substantially alter the $K^-p$ and $K^-$
nucleus interaction and experiments looking for these properties are most welcome. Some 
electromagnetic reactions appear well suited for these studies.
Application of the chiral unitary approach to the
$K^-p\rightarrow\gamma\Lambda$, $\gamma\Sigma^0$ reactions at threshold has
been carried out in [23] and a fair agreement with experiment is found. In
particular one sees there that the coupled channels are essential to get a good
description of the data, increasing the $K^-p\rightarrow\gamma\Sigma^0$ rate
by about a factor 16 with respect to the Born approximation.

        In a recent paper [24] we propose the $\gamma p\rightarrow
        K^+\Lambda(1405)$ reaction as a means to study the properties of the
        resonance, together with the $\gamma A\rightarrow
        K^+\Lambda(1405) A'$ reaction to see the modification of its properties
        in nuclei. The resonance $\Lambda(1405)$ is seen in its decay products
        in the $\pi\Sigma$ channel, but as shown in [24] the sum of the cross
        sections for $\pi^0\Sigma^0$, $\pi^+\Sigma^-$, $\pi^-\Sigma^+$
        production has the shape of the resonance $\Lambda(1405)$ in the I = 0
        channel. Hence, the detection of the $K^+$ in the elementary reaction,
         looking at $d\sigma/dM_I$ ($M_I$ the invariant mass of the meson
         baryon system which can be induced from the $K^+$ momentum), is
         sufficient to get a clear $\Lambda(1045)$ signal. In nuclear targets
         Fermi motion blurs this simple procedure (just detecting the $K^+$), but the
         resonance properties can be reconstructed by observing the decay
         products in the $\pi\Sigma$ channel. In fig. 3 we show the cross
         sections predicted for the $\gamma p\rightarrow K^+ \Lambda(1405)$
         reaction looking at $K^+\pi^0\Sigma^0$, $K^+$ $all$ and $K^+ \Lambda(1405)$
         (alone). All of them have approximately the same shape and strength
         given
         the fact that the I = 1 contribution is rather small.
\vspace{0.3cm}
\begin{figure}[h]
\centerline{\protect
\hbox{
\psfig{file=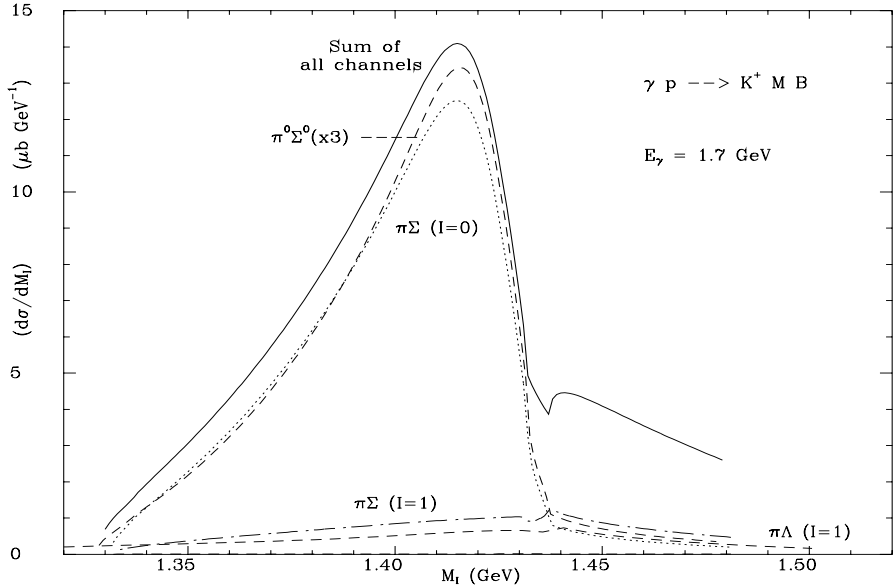,height=8cm,width=13cm,angle=-90}}}
\vspace{0.5cm}
\caption{Cross section for $\gamma p\rightarrow K^+ X$ with $X=all$,
$\pi^0\Sigma^0$, $\Lambda$(1405).}
\end{figure}
\vspace{0.5cm}


        The energy chosen for the photon is $E_\gamma$ = 1.7 GeV which makes it
        suitable of experimentation at SPring8/RCNP, where the experiment is
        planned [15], and TJNAF.
        
        One variant of this reaction is its time reversal $K^- p
\rightarrow\Lambda(1405)\gamma$. This reaction, for a $K^-$ momentum
in the 300 to 500 MeV/c range, shows clearly the $\Lambda(1405)$ resonant production
[25] and
has the advantage that the analogous reaction in nuclei still allows the
observation of the $\Lambda(1405)$ resonance with the mere detection of the
photon, the Fermi motion effects being far more moderate than in the case of
the $\gamma A\rightarrow K^+\Lambda(1405)X$ reaction which requires larger
photon momenta and induces a broad distribution of $M_I$ for a given $K^+$
 momentum.

 One of the interesting developments around these lines is the interaction of
 the $K^-$ with the nuclei, with its relationship to problems like
 $K^-$ atoms or the possible condensation of $K^-$ in neutron stars. The
 problem has been looked at from the chiral perspective by evaluating Pauli
 blocking effects on the nucleons of the intermediate $\bar{K}N$ states
 [26,27].
  These effects lead to a $K^-$ self-energy in nuclei which is attractive
  already at very low densities, as a consequence of pushing the resonance at
  energies above $K^-p$ threshold. However, more recent investigations
  considering the $\bar{K}$ self-energy in a self-consistent way [28] lead to
  quite different results since the resonance barely changes its position. Yet
  one still gets an attractive self-energy which is demanded by the $K^-$ atom
  data [29]. A step forward in this direction is given in [30], where in
  addition to the $K^-$ self-energy in the medium, one also renormalizes the
  pions and takes into account the different binding of N, $\Sigma$ and
  $\Lambda$ in nuclei. Preliminary results from [31] indicate that the 
  $K^-$ self-energy obtained in [30] can lead to a  good microscopical
  description of present data on $K^-$ atoms, hence providing an accurate tool
  to study the properties of $K^-$ at higher densities and the eventual
  condensation in neutron stars.
 \vspace{0.2cm}

\section{ Summary}

  We have reported on the unitary approach to meson meson and meson baryon
  interaction using chiral Lagrangians, which has proved to be an efficient
  method to extend the information contained in these Lagrangians to higher
  energies where $\chi PT$ cannot be used. This new approach has opened the
  doors to the investigation of many problems so far intractable with $\chi PT$
  and a few examples have been reported here. It is clear that these are
  only a few of the many and interesting problems which can now be tackled
  from this perspective. At the same time we have shown that many interesting
  predictions can be tested with present machines, and a few experiments in
  this direction are already planned. Further research, theoretical and experimental, along these
  lines appears to us a very interesting task to undertake in a near future.
\vspace{0.2cm}

\section*{Acknowledgments.}
\vspace{0.1cm}
        
        We are thankful to the COE Professorship program of Monbusho which enabled
        E. O. to stay at RCNP where part of the work reported here has been
        done. E. M. and J. C. N. would like to acknowledge the hospitality
        of the RCNP of the Osaka University and support from the Ministerio de
        Educacion y Cultura. This work is partly supported by DGICYT, contract
        number PB 96-0753.

\end{document}